\documentclass[12pt]{article}

\newcommand{\be}{\begin{equation}}
\newcommand{\bea}{\begin{eqnarray}}
\newcommand{\ee}{\end{equation}}
\newcommand{\eea}{\end{eqnarray}}

\renewcommand{\b}{B^0}
\newcommand{\bb}{\bar {B^0}}

\newcommand{\eq}[1]{Eq.\,(\ref{#1})}

\newcommand{\Dt}{\Delta t}

\newcommand{\Dm}{\Delta m}
\newcommand{\DG}{\Delta \Gamma}
\newcommand{\G}{\Gamma}

\begin{document}

\begin{center}
\bigskip
{\Large \bf \boldmath Direct test of time reversal invariance
violation in B mesons} \\
\bigskip
\bigskip

{\large  Ezequiel \'Alvarez $^{(a)}$\footnote{sequi@df.uba.ar} and
Alejandro Szynkman $^{(b)}$\footnote{szynkman@lps.umontreal.ca} }
\end{center}

\begin{flushleft}
~~~~~~~~~~~$a$: {\it Departamento de F\'{\i}sica Juan Jos\'e Giambiagi,}\\
~~~~~~~~~~~~~~~{\it Facultad de Ciencias Exactas y Naturales,}\\
~~~~~~~~~~~~~~~{\it Universidad de Buenos Aires, Ciudad Universitaria,}\\
~~~~~~~~~~~~~~~{\it Pabell\'on 1, 1428, Buenos Aires, Argentina}\\  
~~~~~~~~~~~$b$: {\it Physique des Particules, Universit\'e de Montr\'eal,}\\
~~~~~~~~~~~~~~{\it C.P. 6128, succ. centre-ville, Montr\'eal, QC, 
Canada H3C 3J7}\\
\end{flushleft}

\begin{center}
\bigskip 
\vskip0.5cm {\Large Abstract\\} \vskip3truemm

\parbox[t]{\textwidth}{{ In this letter we reinterpret and reanalyze
the available data of the B meson factories showing the existence of
direct experimental evidence of time reversal invariance violation in
B mesons. This reinterpretation consists of using the available
observables to define a new observable which, in a model independent
way and without assuming CPT invariance, compares a transition between
a $B^0$ and a here-defined $B_\alpha$-state, with its time reversed
transition. The observable then offers a direct way to probe time
reversal invariance and it is therefore independent of any conclusion obtained from current experimental information on CP violation
and CPT invariance.  As far as the authors are concerned, this is the
first direct evidence of time reversal invariance violation in B
mesons and also the first one obtained from decaying particles whose
mean life time difference is negligible.}}
\end{center}
\thispagestyle{empty}
\newpage
\setcounter{page}{1}
\baselineskip=14pt

In the last fifty years there has been a great theoretical and
experimental advance in the study of the discrete symmetries C (charge
conjugation), P (parity), T (time reversal) and their relevant
combinations CP, T and CPT.  This was achieved thanks to the
exploration of the Electroweak sector of the Standard Model (SM)
where, at the level of energies explored insofar, we understand that
the root of CP and T violation lies.  On the other hand, CPT symmetry
is protected by the general Pauli's CPT theorem and it has not been
measured to be violated. During this period we have learned that the
study of the discrete symmetries may give us clues to project the
behaviour of physics at higher energies. Such is the case of the
celebrated prediction of the third generation~\cite{km} in answer to
the CP violation measured  in the Kaon system~\cite{cronin}.  It is
therefore of great interest the study of the discrete symmetries in
the different sectors of the Electroweak Lagrangian.

The violation of CP has been deeply and extensively studied in the
Kaon system since its first measurement in 1964.  The following
experimental and theoretical studies, in the 70's and 80's, pointed
out that an important CP violation should be expected in the B sector.
In fact, this was measured for the first time by Babar and Belle in
2001~\cite{Aubert,cpv in b}, and since then an amazing amount of data and a
very high accuracy in the measurements have considerably increased our
knowledge of the Electroweak sector.

On the contrary, there is a poor knowledge of T violation, which is a
much more difficult symmetry to be tested.  In fact, it was not
directly measured until 1998 by the CPLEAR
collaboration~\cite{cplear}.  This measurement was performed in Kaon
mesons and, up to now, it is the only direct measurement for T
violation that exists.  This T-violating observable, however, has been
subject of controversy \cite{Wolfy,and friends,pdg} due to its dependence
on the Kaons mean life time difference: since T violation is probed in
this case through the oscillation probability from $K^0 $ to $\bar
K^0$ relative to $\bar K^0$ to $K^0$, it might be that a $\bar K^0$
decaying more rapidly than a $K^0$ mimics a T-violating effect.  The
search for a similar T-violating observable in the B-sector by the
B-factories, which compares the $\Gamma(B^0 \to \bar B ^0,\Dt)$ and
$\Gamma(\bar B ^0 \to B^0,\Dt)$ transitions, it has thrown negative
results~\cite{tv in b,Nakano}, as expected due to the negligible mean life
time difference ($\DG$) in the neutral B mesons. As a matter of fact,
T-invariance is also proposed to be explored in the B
system through T-odd observables~\cite{3p} which show up through the
angular analysis of certain $B$ decays.

The present state of the art indicates that T violation should be
expected in the B system. However, owing to its importance, it is of
great interest to define and measure a T-violating observable which
is, in addition, not null in the limit of a vanishing $\DG$. In fact,
in this case the resulting observable will lack of the previously
mentioned controversy and, hence, it can be taken as the first time
reversal violation effect in the sense of
Ref.~\cite{Wolfy}. Therefore, the goals of this work are the
following: $(i)$ To show how a proposed T-violating observable for the
B-factories exhibits, in a model independent way and without assuming
CPT invariance, direct evidence of T violation in a transition between
some given neutral B-states; and, therefore, it is 
independent of any indirect evidence of T violation obtained from our
present knowledge on CP violation and CPT invariance in B mesons.  And
$(ii)$ to reanalyze the available experimental data and compute this
T-asymmetry to positively find direct evidence of T violation in the B
system.
 
The observable that we propose in this letter has been previously
studied in Refs.~\cite{seq,Alvarez,Banuls,Banuls2}.  However, in those
articles, the interpretation of this observable as a T-asymmetry is
only valid within the SM, where the formalism of the CP-tag can be
applied. Along this work, we go beyond this formalism and we construct
the T-violating observable only from quantum mechanics and available
experimental information. In addition, we also compute its value from
current data.

The T-operation, in contrast with the CP-operation, corresponds to an
antiunitary operator and hence it has no conserved quantum numbers.
$\hspace*{-0.1cm}$ Therefore, we cannot assert that T is either
conserved or not in a given process or decay; instead, we should say
that the process is {\it T-invariant} or not.  Hence, a way to
inquire directly about the T-invariance of a given process is by
comparing it with its T-reversed. For instance, if $\phi_1$ and
$\phi_2$ are two possible quantum states, then the inequality \bea
|\langle \phi_2 | U(\Dt) | \phi_1 \rangle|^2 \neq |\langle \phi_1 |
U(\Dt) | \phi_2 \rangle|^2 ,
\label{te viole}
\eea
where $U(\Dt)$ stands for the evolution operator, is a clear signal that the process is not T-invariant (usually referred as {\it T violation}, for short). A direct measurement of an inequality of this type is an explicit measurement of T violation, as it was the case of CPLEAR in the Kaon system~\cite{cplear}.  We show here that in the B-factories it is possible to construct an asymmetry which probes, in a model independent way and without assuming CPT-invariance, an inequality of the type in \eq{te viole} and, hence, it offers a direct test of T violation in the B system. Moreover, since $\DG$ is negligible in the B system, this asymmetry is free from any argument that might arise due to a considerable mean life time difference.

For the purposes of this letter, we consider the events in the B-factories in which the $\Upsilon(4S)$ decays to a pair of correlated neutral B mesons. Once this correlated pair is created, the first decay of one of these mesons tags the other meson, which evolves during a $\Dt$ time until it decays.   The process of {\it tagging} in a correlated couple of particles occurs always no matter which is the first decay. In the B-factories, where the correlated initial state at time $t$ is written as
\bea
|i(t)\rangle = 
\frac{e^{-\G t}}{\sqrt{2}} \left( | \b (+\vec k),\bb(-\vec k)\rangle - |\bb(+\vec k), \b(-\vec k)\rangle \right) ,
\label{i}
\eea
a first decay to $X$ at time $t_1$ projects the second meson which is still flying to $$e^{-\G t_1} |B_{\bar X} \rangle \equiv  \frac{1}{\sqrt{2}}e^{-\G t_1}(\langle X |T| \b \rangle |\bb\rangle - \langle X |T|\bb \rangle | \b\rangle ),$$
where $\Gamma^{-1}$ is the mean life time of the B mesons and $T$ is the 'scattering' matrix of the direct decay  --with no mixing--.  Hence, we say that the first decay has acted as a filter in the quantum mechanical sense and has tagged the second meson as a $|B_{\bar X}\rangle$.  The state of the second meson is as physical and determined as it is --for instance-- a flavour state $B^0$, even if the $B \to X$ decay is not theoretically well understood.  Therefore, every time that we have an $X$ decay on one side, we know that a $B_{\bar X}$ must occur on the other side, and this conclusion is achieved beyond any 'a priori' theoretical assumption regarding the decay process. 
%%%%%%%%%%%%%%%%%%%%%%%%%%%% 

Suppose that the first decay of the correlated B pair is $B\to X = J/\psi K_S$, then the tagged meson is 

\bea
| B_{\overline{J/\psi K_S}} \, \rangle &=&  \frac{1}{\sqrt{2}} (\langle K_S |T| \b \rangle |\bb\rangle - \langle K_S |T| \bb \rangle | \b\rangle ) . \nonumber
\eea
(Here and below  $|K_{S,L} \rangle$ stands for $|J/\psi K_{S,L}\rangle$). This motivates us to define a new orthonormal basis in the B-space such that one of its vectors is parallel to $B_{\overline{J/\psi K_S}}$ ,
\bea
|B_\alpha \rangle &=&  \frac{1}{\sqrt{N}} (\langle K_S |T| \b \rangle |\bb\rangle - \langle K_S |T| \bb \rangle | \b\rangle  ) \nonumber \\
| B_{\alpha_\perp} \rangle &=& \frac{1}{\sqrt{N}} \left( \langle K_S |T| \bb \rangle^*\,  |\bb\rangle + \langle K_S |T| \b \rangle^* \,|\b \rangle \right) , \nonumber
\eea
where $N$ is the normalization factor.  These two new states are well and unambiguously defined through these equations, and hence they are physical.  It is worth noticing at this point that due to Bose-statistics we can assert a first important feature of this basis, namely the impossibility of $B_\alpha$ to decay directly to $J/\psi K_S$:
\bea
\langle K_S |T|B_\alpha\rangle &=& 0 .
\label{salvadora}
\eea

Using this new basis, we can rewrite the initial state of the B-factories at time $t$, \eq{i}, as  
\bea
|i(t)\rangle =  
\frac{e^{-\G t}}{\sqrt{2}} \left( | B_\alpha (+\vec k), B_{\alpha_\perp} (-\vec k)\rangle - |B_{\alpha_\perp} (+\vec k), B_\alpha (-\vec k)\rangle \right) \nonumber .
\eea
This new expression allows us to understand, with help of \eq{salvadora}, that the $B_{\alpha_\perp}$ state is the {\it father} of the $J/\psi K_S$ decay.
%%%%%%%%%%%%%%%%%%%%%%%%%%%

We are now interested in showing that $B_\alpha$ is the {\it father}
of the $J/\psi K_L$ decay and that its branching fraction is equal to
that corresponding to the decay $B_{\alpha_\perp}\to J/\psi K_S$. Up
to here we have used essentially quantum mechanics in our derivation
but, in order to achieve this goal, we also need the input of
experimental information.  In this direction, we first obtain the
following relation from measurements of direct CP violation in $B \to
J/\Psi K_{S,L}$ decays~\cite{belle-nuevo} \footnote{Although
contrived, these measurements may be also interpreted as a fine tuned
NP cancellation between CP-violation in the mixing and direct
CP-violation in the decay in the $B \to J/\psi K_{S,L}$ channels.
However, this possibility is unrealistic since it should be also
consistent with the measurement of the semileptonic asymmetry
(see~\cite{tv in b,Nakano} and footnote [5]).}
 
\bea
|\langle K_{S,L} |T| B^0 \rangle   | &=& |\langle K_{S,L} |T| \bar B^0  \rangle | \hspace*{0.1cm} ( 1 + \eta_1) ,
\label{exp result}
\eea   
with $\eta_1$ being at most of a few percent
($\eta_1$$\sim$$10^{-2}$). Next, we express the $K_{S,L}$ states in
the $\{K^0,\bar K^0\}$ flavour basis (a small parameter,
$\eta_2$$\sim$$10^{-3}$, accounts for Kaons CP-violation in this
change of basis). Then, neglecting $\eta_{1,2}$ (the
$\eta_{1,2}$$\neq$$0$ case is analyzed below) and combining \eq{exp
result} --now written in terms of $K^0$ and $\bar K^0$-- with $\langle
\bar K ^0 |T|B^0\rangle=\langle K ^0 |T|\bar B ^0
\rangle$=0~\footnote{A small departure from the equality $\langle \bar
K ^0 |T|B^0\rangle = \langle K ^0 |T|\bar B ^0 \rangle = 0$ would be
still acceptable. In any case, these transitions would involve
processes which are highly constrained by, for example, $\Delta S = 1$
and $\Delta B = 1$ weak neutral currents modes in $K^0$ and $B^0$
decays respectively (see~\cite{pdg2}).}, we obtain a $T$-matrix
proportional to the identity in the $K$-flavour and $B$-flavour
basis. Afterwards, we perform two unitary rotations to the $\{K_S,K_L
\}$ and $\{ B_{\alpha},B_{\alpha_\perp} \}$ basis and we use
\eq{salvadora} to get
\bea  \langle K_L | T | B_{\alpha_\perp} \rangle
&=& 0
\label{marado}
\eea
and 
\bea 
|\langle K_S | T | B_{\alpha_\perp} \rangle | &=& |\langle K_L | T | B_{\alpha} \rangle |,
\label{garcha}
\eea
as we wanted to show.

The result in \eq{marado} is of great usefulness since, together with \eq{salvadora}, it allows us to reduce to a {\it single transition} between well defined B-states the following two intensities:
\bea
I(\ell^-,K_L,\Dt) = c\, |\langle K_L |T|B_{\alpha}\rangle|^2 |A_{\ell^-}|^2 |\langle B_\alpha | U(\Dt) | B^0 \rangle |^2 \hspace*{0.25cm}
\label {caca} \\ 
I(K_S,\ell^+,\Dt) = c\, |\langle K_S |T|B_{\alpha_\perp}\rangle|^2 |A_{\ell^+}|^2 |\langle B^0 | U(\Dt) | B_{\alpha} \rangle |^2 , 
\label {pipi} 
\eea
where $I(X,Y,\Dt)$ is the probability of having first an $X$ decay and $\Dt$ later an $Y$ decay, $\ell^\pm$ refers to flavour specific leptonic decays, $|A_{\ell^\pm}|^2$ is their corresponding branching ratios and '$c$' is a common constant factor to both intensities.  
%%%%%%%%%%%%%%%%%%%%%%%

Motivated by the time reversed single B-meson transitions that occur in Eqs.~(\ref{caca}) and (\ref{pipi}), we propose to measure the following asymmetry,
\vspace*{-0.1cm}
\bea
A^{exp}_T (\Dt) &=& \frac{ I(\ell^-,K_L,\Dt) - I(K_S,\ell^+,\Dt) }{ I(\ell^-,K_L,\Dt) + I(K_S,\ell^+,\Dt) } .
\label{alapelotita}
\eea 
As it is easily seen using \eq{garcha} and assuming
$|A_{\ell^+}|=|A_{\ell^-}|$~\footnote{Observe that this equality is a
sufficient but not a necessary condition for our argument, and only a
deviation of order 10$\%$ would spoil our conlusions in \eq{kk}.  This
--widely accepted-- equality has not been directly measured, since it
is very difficult to separate it from indirect CP-violation in an
observable.  However, the precise measurement of the semileptonic
asymmetry~\cite{tv in b,Nakano} rules out a departure greater than the
$1\%$ level from the above equality, unless a fine tuned new physics
cancellation between a deviation of this equality and indirect
CP-violation occurs.  However, in close connection with the discussion
given in footnote [3], such fine tuning is highly unrealistic
since it should be also consistent with the measurements of the
combination of CP violation in mixing and in decay --${\cal A}_f$ in
the notation of the paper quoted in~\cite{belle-nuevo}-- from
time-dependent decay rates in non leptonic B decays (see, for
instance, the article given in~\cite{belle-nuevo}).}, this asymmetry
gets reduced to a T-violating asymmetry in the spirit of \eq{te
viole}, 
\bea \left. A^{exp}_T \right|_{\eta_1=\eta_2=0} = A_T (\Dt)
\equiv \frac{|\langle B_{\alpha} | U(\Dt) | \b \rangle |^2 - |\langle
\b | U(\Dt) | B_\alpha \rangle|^2 }{|\langle B_{\alpha} | U(\Dt) | \b
\rangle |^2 + |\langle \b | U(\Dt) | B_\alpha \rangle|^2 } .
\label{pampita}
\eea
This expression represents an asymmetry constructed with a well defined transition between physical states and its T-reversed process without assuming CPT-invariance, the\-re\-fo\-re it is an asymmetry sensitive to T violation in a direct way and then independent of any deduction which combines measurements of CP violation and CPT invariance. 

The case $\eta_{1,2}\neq 0$ is now easily analyzed perturbatively.  In fact, if we think of $\eta$ as a small parameter whose magnitude is at most as large as the maximum between $\eta_1$ and $\eta_2$, then we have that a term of order $\eta$ should be added in both RHS of Eqs.~(\ref{marado}) and (\ref{garcha}).  These extra terms end up giving a correction to \eq{pampita}, which now reads 
\bea
A^{exp}_T (\Dt) = A_T (\Dt) + {\cal O}(\eta) . \nonumber
\label{pampita2}
\eea
Therefore, the conclusion is straightforward: {\it the measurement of
\bea
|A^{exp}_T (\Dt)| \gg \eta \sim 10^{-2}
\label{kk}
\eea
for some $\Dt$, implies direct evidence of time reversal invariance violation in the mixing of the B mesons in a model independent way.} 

The conclusion concerning \eq{kk} is now easily analyzed using the available experimental data of Babar and Belle. We stress at his point that, as far as the authors are concerned, the particular combination of intensities which gives rise to the T-asymmetry $A^{exp}_T$, \eq{alapelotita}, has not been computed in the literature. The experimental results in References \cite{belle-nuevo} and \cite{babar-nuevo} related to $I(K_S,\ell^+,\Dt)$ and  $I(\ell^-,K_L, \Dt)$ are obtained through a fit to a formula which assumes $\Delta\Gamma=0$ (this condition assures that our observable is a true T-violating observable in the spirit of reference \cite{Wolfy}). Since current experiments~\cite{tv in b,Nakano} constrain $\Delta\Gamma/\Gamma\leq \nolinebreak 10^{-2}$, we can safely use this fit: any correction to our results would be at most of this order of magnitude and, hence, it will not modify our conclusion. 

The result for the time dependence of the relevant intensities is
\bea
I(\ell^-,K_L,\Dt) &\propto&
\frac{e^{-\G\Dt}}{4\Gamma} \times
\left\{
\begin{array}{ll}
\left( 1 + (0.716 \pm 0.080) \sin(\Dm\Dt) \right) \hspace*{0.2cm} \mbox{Babar~\cite{babar-nuevo}} \\
\left( 1 + (0.641 \pm 0.057) \sin(\Dm\Dt) \right) \hspace*{0.2cm} \mbox{Belle~\cite{belle-nuevo}} 
\end{array}
\right. \nonumber 
\eea
\vspace*{-0.5cm}
\bea
I(K_S,\ell^+,\Dt) &\propto& \nonumber 
\frac{e^{-\G\Dt}}{4\Gamma} \times
\left\{
\begin{array}{ll}
\left( 1- (0.691 \pm 0.040) \sin(\Dm\Dt) \right) \hspace*{0.2cm} \mbox{Babar~\cite{babar-nuevo}}\\
\left( 1- (0.643 \pm 0.038) \sin(\Dm\Dt) \right) \hspace*{0.2cm} \mbox{Belle~\cite{belle-nuevo}} 
\end{array}
\right. 
\nonumber
\eea
and therefore, using \eq{alapelotita}, we obtain the result for the model independent T-violating observable,
\bea
A^{exp}_T (\Dt) = 
\left\{
\begin{array}{ll}
(0.703 \pm 0.044) \sin(\Dm\Dt)  & \mbox{Babar} \\
(0.642 \pm 0.034) \sin(\Dm\Dt)  & \mbox{Belle.} \quad
\end{array}
\right.
\label{platino}
\eea
As it is easily seen, \eq{platino} clearly 
satisfies the requirement in \eq{kk} and, therefore, we can
positively assert that a reinterpretation and a reanalysis of the
B-factories available data provides a direct evidence
of T violation in the B system.  

Even if T violation should be expected in this sector, the relevance
of the result resides in defining properly --i.e., without assuming
CPT-invariance and model independently-- a suitable observable to
probe it and, finally, in obtaining a direct experimental evidence of
T violation.  It is worth to stress that the T-violating
observable is, in consequence, independent of any result arising from
constrains imposed by measurements of CP violation and CPT
invariance.

In summary, we have reanalyzed the observables measured by the
B-factories and we have shown that they imply direct evidence of T
violation in the B system, where the negligible mean life time
difference reinforces this result. We have proposed an
asymmetry which, in a model independent way and without assuming CPT
invariance, consists of the comparison of the transition between a
$B^0$ and a $B_\alpha$-state, and its T-reversed: $\G(B^0\to
B_\alpha,\Dt)$ versus $\G(B_\alpha \to B^0,\Dt)$.  We have shown how
the precise definition of the $B_\alpha$-state arises from the concept
of tagging, which filters the information of a given decay into the
structure of the tagged meson.  The argument in this letter states
that if the measurement of the proposed T-asymmetry, $A^{exp}_T(\Dt)$
(\eq{alapelotita}), is greater than the quoted $\eta\sim 10^{-2}$ for
some $\Dt$ then there is a direct signal of T violation.  The
reanalysis of the available experimental data fully overshoots this
condition by more than ten standard deviations (\eq{platino}) and
hence it provides clear direct evidence of T violation in the B
system. 

We are grateful with J.~Bernab\'eu, who encouraged us to write this paper, for being part in the germination of the idea and for very helpful discussions. We also wish to thank D.~London and L.~Wolfenstein for very helpful communications during the development of this work; and S.~Brunet, Y.~Sakai and C.~Yeche for very fruitful experimental discussions.

% y que se cague la choca toreando !!!!

\end{document}